\documentclass[11pt, a4paper, Physsubmission, Phys]{SciPost} % forces arXiV to use A4 format
\hypersetup{
    colorlinks,
    linkcolor={red!50!black},
    citecolor={blue!50!black},
    urlcolor={blue!80!black}
}

\usepackage{indentfirst}
\usepackage[bitstream-charter]{mathdesign}
\DeclareSymbolFont{usualmathcal}{OMS}{cmsy}{m}{n}
\DeclareSymbolFontAlphabet{\mathcal}{usualmathcal}

\begin{document}

\begin{center}{\Large \textbf{
On irregularities in the cosmic ray spectrum of \(10^{16}-10^{18}\) eV range\\
}}\end{center}

% TODO: write the author list here. Use initials + surname format.
% Separate subsequent authors by a comma, omit comma at the end of the list.
% Mark the corresponding author with a superscript *.
\begin{center}
S.\,P. Knurenko and
I.\,S. Petrov\textsuperscript{$\star$}
\end{center}

\begin{center}
Yu. G. Shafer Institute of Cosmophysical Research and Aeronomy, Yakutsk, Russia
\\
* igor.petrov@mail.ysn.ru
\end{center}

\begin{center}
\today
\end{center}

% For convenience during refereeing (optional),
% you can turn on line numbers by uncommenting the next line:
%\linenumbers

\definecolor{palegray}{gray}{0.95}
\begin{center}
\colorbox{palegray}{
  \begin{tabular}{rr}
  \begin{minipage}{0.1\textwidth}
    \includegraphics[width=30mm]{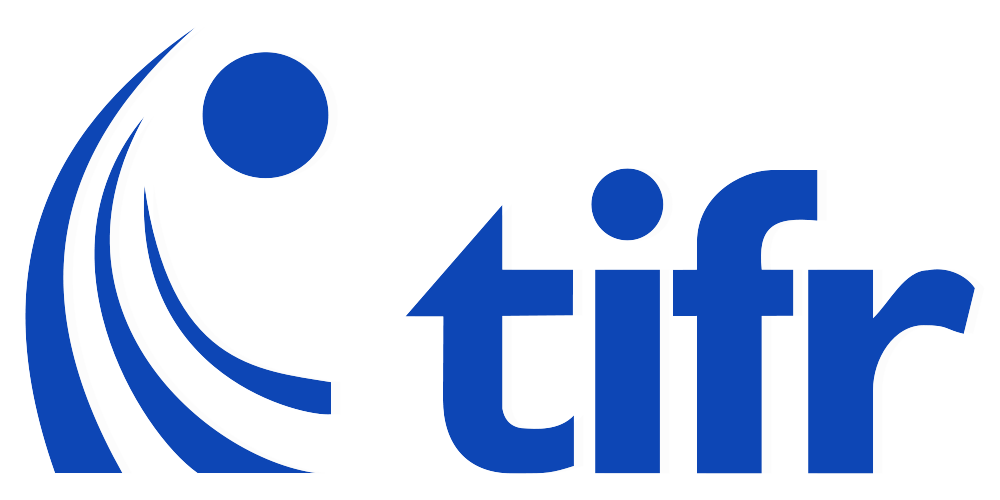}
  \end{minipage}
  &
  \begin{minipage}{0.85\textwidth}
    \begin{center}
    {\it 21st International Symposium on Very High Energy Cosmic Ray Interactions (ISVHE- CRI 2022)}\\
    {\it Online, 23-27 May 2022} \\
    \doi{10.21468/SciPostPhysProc.?}\\
    \end{center}
  \end{minipage}
\end{tabular}
}
\end{center}

\section*{Abstract}
{\bf
% TODO: write your abstract here.
Existing small, medium and large arrays for the study of cosmic rays of ultra-high energies are aimed for obtaining information about our galaxy and extragalactic space, namely to search and study astronomical objects that produce the flux of relativistic particles. The drift and interaction of such particles with magnetic fields and shock waves taking place in interstellar space causes the same interest. The shape of the energy spectrum of cosmic rays in the energy range \(10^{15}-10^{18}\) eV, where the "knee" and the "second knee" are observed, can be formed as a superposition of the partial spectra of various chemical elements. Verification of galactic models, using recent experimental spectral data, makes it possible to study the nature of the galactic and extragalactic components of cosmic rays. The paper presents the result of the energy spectrum of cosmic rays in the range \(10^{16}-10^{18}\) eV of measurements obtained with the Small Cherenkov array --- a part of the Yakutsk array.
}

\section{Introduction}
\label{sec:intro}
% TODO: write your article here.
The study of ultra-high energy cosmic rays has the following goals: determination of anisotropy, energy spectrum and mass composition. These are important to advance our understanding of the origin, acceleration and propagation of cosmic rays of different energies. There are different methods for different energy ranges: for energies \(10^{12}-10^{14}\) eV --- direct measurements on satellites ~\cite{Adriani20113326972, Aguilar2015114171103, Green2018301159} and on balloons ~\cite{Panov200973564, Seo2008421656} for energies greater than $10^{14}$ eV --- indirect measurements by extensive air showers method, i.e. by tracking the cascade processes in the atmosphere and detecting charged particles fluxes. Since the spectrum of cosmic rays is very wide, there are experiments with different sizes: compact ones with an area of s < 1 km\(^{2}\) --- air showers up to energies of $10^{18}$ eV, average sized arrays with s < 20 km\(^{2}\) --- air showers up to energies of $10^{19}$ eV and huge arrays such as Auger~\cite{Aab2015798172}, Telescope Array~\cite{AbuZayyad2012689} for even greater energies.

\section{The Small Cherenkov Array}

\begin{figure}[ht]
	\centering
	\includegraphics[width=0.8\textwidth]{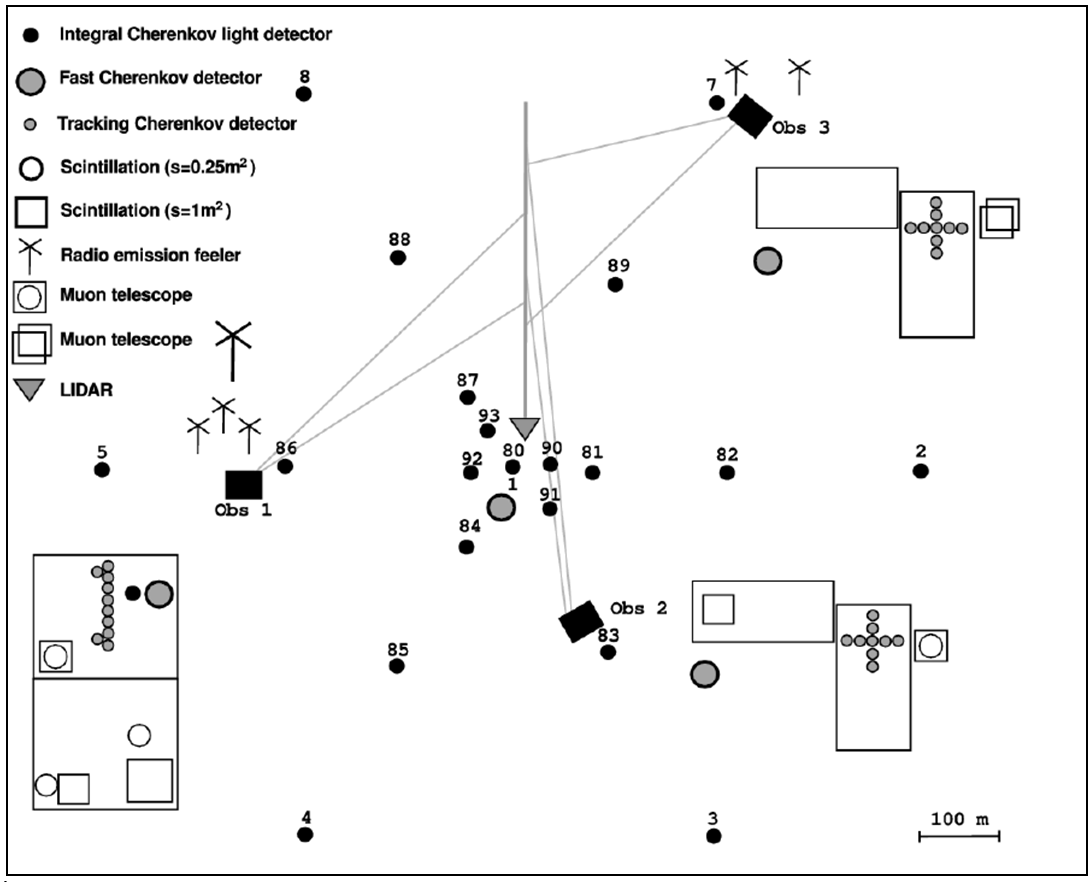}
	\caption{The layout of the detectors of charged particles, muons, and Cherenkov light of the Small Cherenkov array located in the center of the Yakutsk array~\cite{Knurenko201982732}.}\label{fig:ykt_1}
\end{figure}

The Small Cherenkov array is a part of the Yakutsk array with an area of 1 km\(^2\) and is aimed to register air showers with energies \(10^{15}-10^{18}\) eV (Fig.~\ref{fig:ykt_1}). The distinctive feature of the Small Cherenkov is that it measures several components of air showers like muons, electrons and Cherenkov radiation, unlike other compact arrays like  KASCADE~\cite{Apel201236183} or TUNKA~\cite{Prosin201475694}. Such hybrid measurements provide a broader outlook at the development of the shower, including longitudinal development, registering the spatial distribution of electrons, muons and Cherenkov light at sea level~\cite{Knurenko20011157, Knurenko199844650}. The longitudinal profile is measured via the flux of Cherenkov photons by the Cherenkov tracking detectors~\cite{Knurenko199844650, Knurenko200683473}.

\section{Air Shower Measurement Simulation at the Small Cherenkov Array}

The measurement precision was determined by full Monte Carlo simulation. The precision of air shower characteristics is shown in Table~\ref{tab:tab_1}.

\begin{table}[htbp]
	\caption{\(E_0\) --- air shower energy [PeV]; \(\sigma\)(R) --- air shower axis reconstruction error [m]; \(\sigma N_s\) --- total number of charged particles determination error, \(\sigma(Q(100))\), \(\sigma(Q(200))\) and \(\sigma(Q(400))\) --- errors of determining the classification parameters of the Cherenkov radiation flux at a distance 100, 200 and 400 m respectively \([\text{phot.}/\text{m}^2]\); \(\sigma(\rho_s(300))\) and \(\sigma(\rho_s(600))\) --- errors of determining of total charged particles flux density at 300 and 600 m respectively \([1/\text{m}^2]\); \(\sigma(\theta\)) --- zenith angle determination uncertainty \([\,^\circ]\) }
	\label{tab:tab_1}
	\begin{tabular}{|c|c|c|c|c|c|c|c|c|}
		\hline
		\(E_0\) & \(\sigma(R)\) & \(\sigma N_s\) & \(\sigma(Q(100))\) & \(\sigma(Q(200))\) & \(\sigma(Q(400))\) & \(\sigma(\rho_s(300))\) & \(\sigma(\rho_s(600))\) & \(\sigma(\theta\)) \\ \hline
		   2    &      9.7      &      0.15      &        0.17        &         -          &         -          &            -            &            -            &    1.3     \\ \hline
		  10    &      7.2      &      0.11      &        0.15        &         -          &         -          &            -            &            -            &    1.0     \\ \hline
		  100   &     15.5      &      0.27      &        0.15        &        0.25        &         -          &            -            &            -            &    1.7     \\ \hline
		  200   &     14.6      &      0.32      &        0.20        &        0.20        &        0.22        &          0.25           &                         &    1.4     \\ \hline
		 1000   &     16.7      &      0.35      &         -          &         -          &        0.20        &          0.17           &          0.19           &    1.3     \\ \hline
	\end{tabular}	
\end{table}

Measurements are also affected by light loss of aerosol particles in the atmosphere. In addition, the transparency of the atmosphere depends on climate, e.g. the non standard atmosphere formed above the Yakutsk array in winter. As shown in the papers~\cite{Knurenko200683473, Knurenko20149292}, near-ground mists and haze during winter can cause significant attenuation of the Cherenkov light flux up to 30-40 \% at distances of 100-400 m from the shower axis, while under excellent weather conditions losses do not exceed 10\%. At the Yakutsk array, the atmosphere is observed regularly~\cite{Knurenko20149292} and these atmospheric measurements are taken into account when determining characteristics of air showers, in particular, energy and depth of maximum development  ~\cite{Knurenko201710466}.

\begin{figure}[ht]
	\centering
	\includegraphics[width=0.5\textwidth]{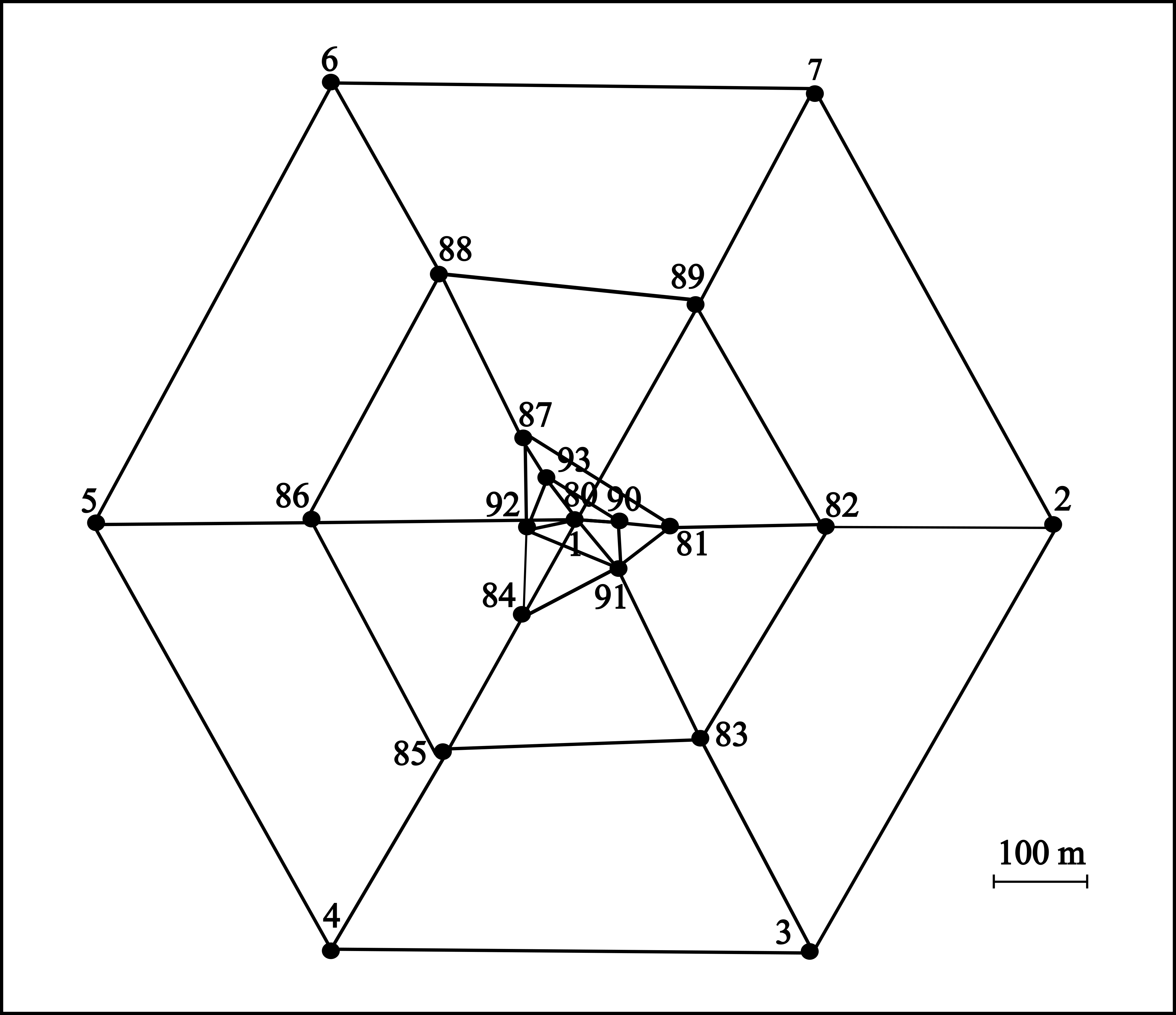}
	\caption{The configuration of trigger triangles used on the Small Cherenkov array~\cite{Knurenko201982732}.}\label{fig:ykt_2}
\end{figure}

In addition, the simulation algorithm included the simulation of the trigger conditions, taking into account the detector's threshold, and fluctuations of the threshold in the conditions of background noise while measuring the flux of Cherenkov light. The Yakutsk array uses two triggers aimed for different energy ranges: \(10^{15}-10^{18}\) eV --- Small Cherenkov array trigger; \(\geq 10^{17}\)  eV --- main trigger from ground scintillation detectors. Fig.~\ref{fig:ykt_2} shows the implementation scheme of the Small Cherenkov array trigger triangles for air showers with energies of \(10^{15}-10^{18}\) eV

Since different triggers select air shower events with different energies, there should be a transition effect --- at the energy boundary, the effective collection area of showers will vary. For large showers, it is underestimated, and for small showers, on the contrary, it is overestimated. With a correction of the effective area for events with high-energy showers, this effect is estimated as 15\%.

\section{Air Shower Energy Estimation}

The accuracy of the estimated number of charged particles at sea level is 10-25\%, and for the total flux of Cherenkov light it is 10-15\% for air showers with energies < 5\(\cdot10^{17}\) eV.  The accuracy is good enough to get the necessary information to obtain the parameters of air showers. The calorimetric method is used for the energy estimation which takes all measured air shower components into account: electrons, muons, and the total Cherenkov light flux. Data were recorded at the Small Cherenkov array for the period 1994 to 2014.

At the Yakutsk array, the primary energy of the particle that produced the air shower is determined with the energy balance method:

\begin{equation}
	E_0 = E_{ei} + E_{el} + E_{\mu} + E_{hi} + E_{\mu i}+ E_{\nu}
\end{equation}

The parameters of formula (1) were determined empirically, for energies in the range \(5\cdot10^{15}-3\cdot 10^{17}\) eV. More information about the method can be found in~\cite{Knurenko200683473}.

The energy estimation with the parameters Q~(100), Q~(200) and Q~(400) has lower uncertainty since the dependence on the zenith angle is weak. We used the following formulas to determine these parameters:

\begin{equation}
	\label{eq:eq_2}
	E_0 = (5.75\pm1.39)\cdot 10^{16}\cdot \left(\frac{Q(100)}{10^{7}}\right)^{(0.96\pm0.03)}
\end{equation}

\begin{equation}
	\label{eq:eq_3}
	E_0 = (1.78\pm0.44)\cdot 10^{17}\cdot \left(\frac{Q(200)}{10^{7}}\right)^{(1.01\pm0.04)}
\end{equation}

\begin{equation}
	\label{eq:eq_4}
	E_0 = (8.91\pm1.96)\cdot 10^{17}\cdot \left(\frac{Q(400)}{10^{7}}\right)^{(1.03\pm0.05)}
\end{equation}

where Q~(100), Q~(200), Q~(400) --- Cherenkov light flux density at the distances 100, 200 and 400 m respectively. 

We can estimate the energy scattered in the atmosphere above the observation level by electrons with the following equation:

\begin{equation}
	\label{eq:eq_5}
	E_{ei} = k(x,P_{\lambda})\cdot\Phi
\end{equation}

where \(\Phi\) is the total flux of Cherenkov light; \(k(x,P_{\lambda})\) --- proportionality coefficient, depends on the transparency of the atmosphere \(P_\lambda\), and the longitudinal development of the shower (energy spectrum of secondary particles and its dependence on the age of the shower), which is expressed through \(X_{max}\), measured at the array.

The energy carried to the observation level by electrons is
\begin{equation}
	\label{eq:eq_6}
	E_{el} = 2.2\cdot 10^{6} \cdot N_s(X_0)\cdot \lambda_{eff}
\end{equation}
where \(N_{s} (X_{0})\) is the total number of charged particles at sea level, and \(\lambda_{eff}\) is the absorption range of the shower particles~\cite{Knurenko200683473}.

\begin{figure}[ht]
	\centering
	\includegraphics[width=0.5\textwidth]{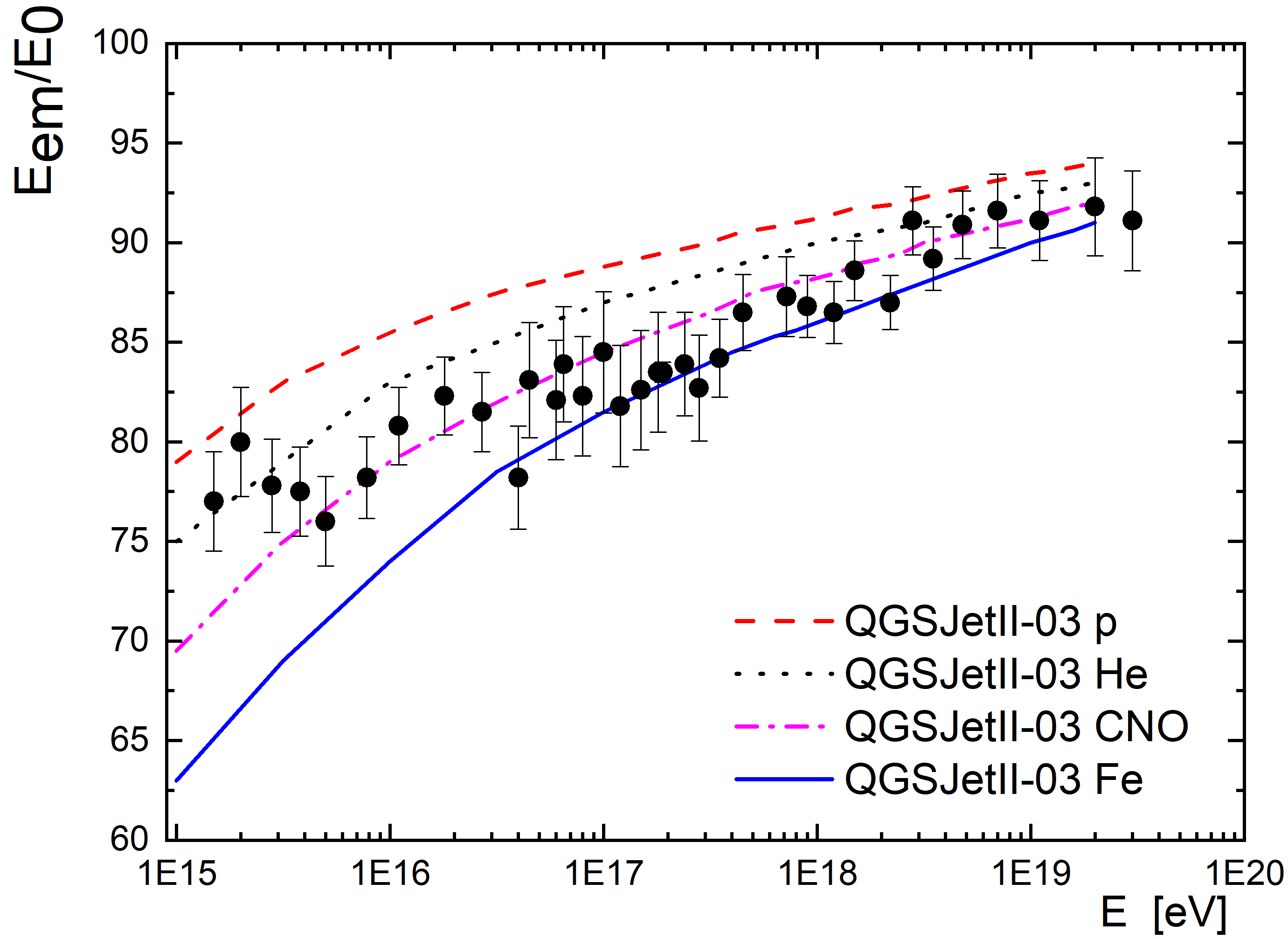}
	\caption{The fraction of energy transferred to the electromagnetic component according to the registration of Cherenkov light at the Yakutsk array and the QGSJetII-03 model of hadron interactions for the proton p (dash), helium He (dots), CNO nuclei (dash-dot) and iron Fe nucleus (solid).}\label{fig:ykt_3}
\end{figure}

Fig.~\ref{fig:ykt_3} shows the dependence of \(E_{em} / E_{0}\) on energy and comparison of the experimental data with QGSJetII-03~\cite{Ostapchenko2006151} simulations taken from~\cite{Abbasi201886574}. The parameter \(E_{em}\) --- is the energy transferred to the electromagnetic component of the air shower and is equal to \(E_{em} = E_{ei} + E_{el}\), and \(E_0\) --- total air shower energy. 

As can be seen from Fig.~\ref{fig:ykt_3} on average, the experimental data are consistent with calculations based on the QGSjet-03 model of hadron interactions and a mixed composition of primary particles.

\section{Energy spectrum of air showers}

\begin{figure}[h!]
\begin{minipage}[h]{0.8\linewidth}
	\center{\includegraphics[width=0.8\textwidth]{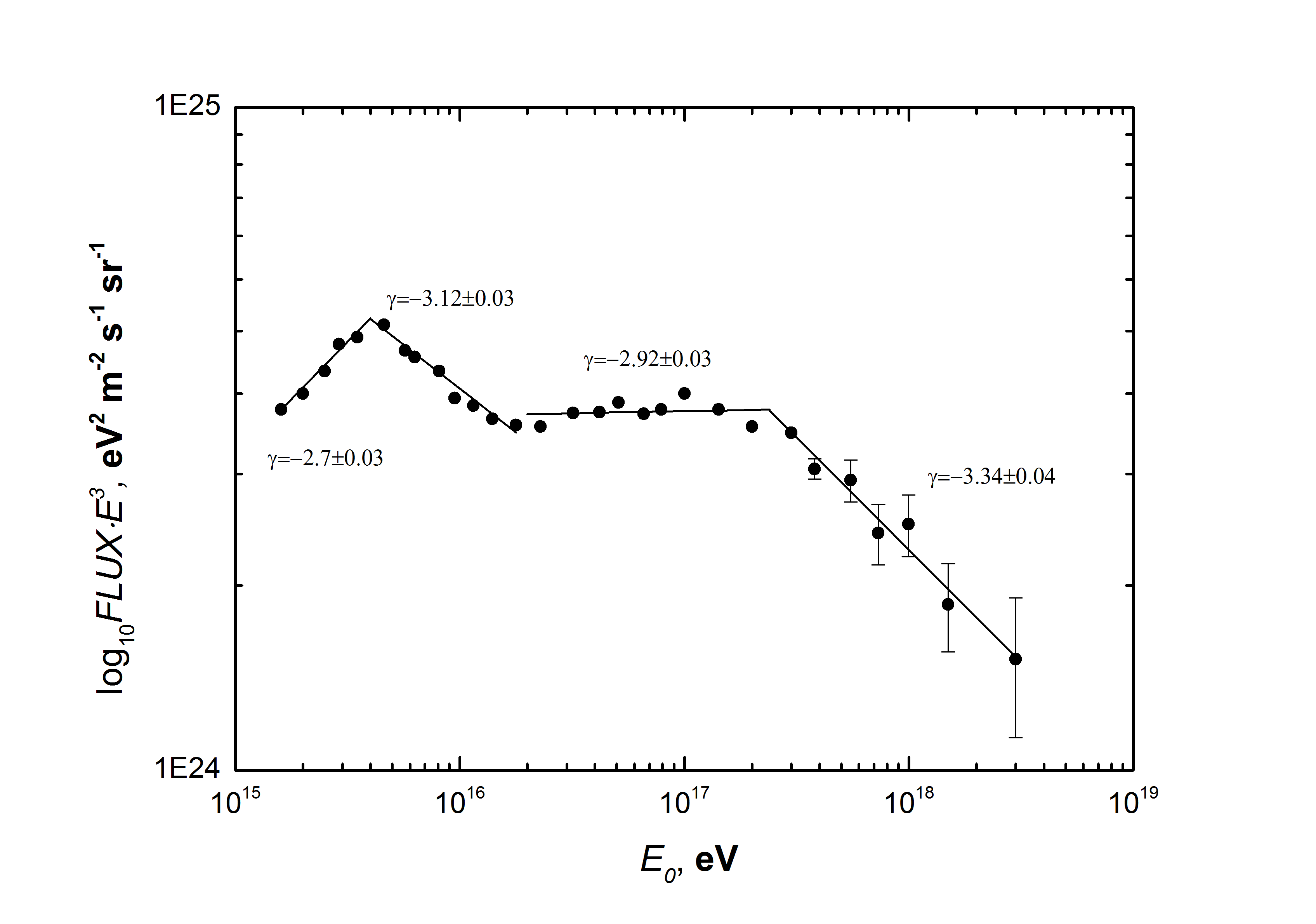}\\a)}	
\end{minipage}
\hfill
\begin{minipage}[h]{0.8\linewidth}
	\center{\includegraphics[width=0.8\textwidth]{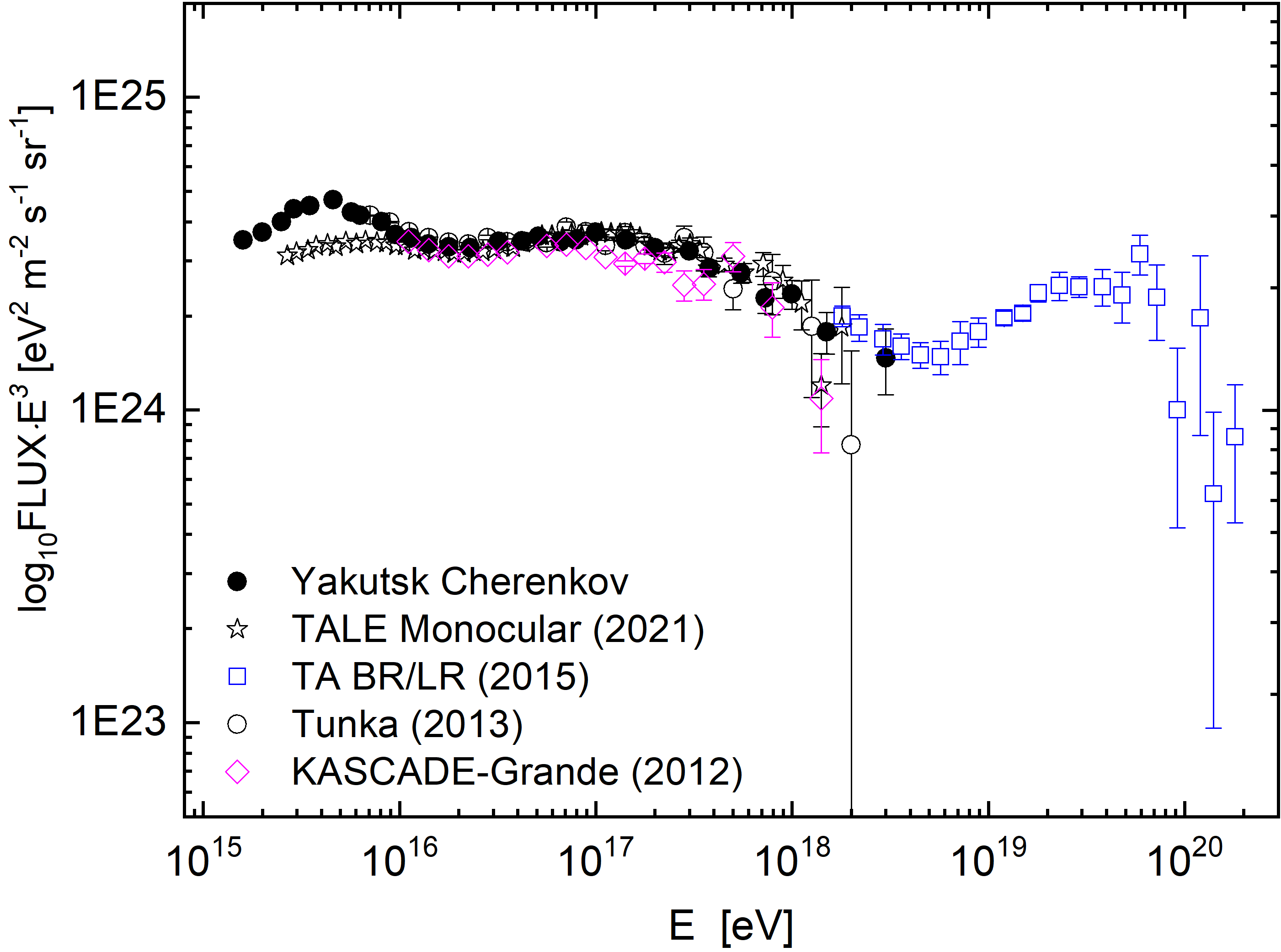}\\b)}	
\end{minipage}
\caption{a) Cosmic ray spectrum in the region \(10^{16}-10^{18}\) eV by Yakutsk data~\cite{Knurenko201982732};
		 b) Comparison of the spectra of the Yakutsk array (dots), TALE~\cite{Abu-Zayyad2022395347} (stars), TA-BR/LR (squares)~\cite{Abbasi201680}, KASCADE-Grande (diamonds)~\cite{Apel201236183} and Tunka (circles)~\cite{Prosin201475694}}
\label{fig:ykt_45}
\end{figure}

Using data of the Small Cherenkov array from 1994 to 2014, we estimated the intensity of air showers in a given intervals of energy \(\Delta E_{i}\) and the zenith angle \(\Delta\theta_{i} \) per unit of the effective area of the array. Fig.~\ref{fig:ykt_45}a shows the resulting spectrum for \(10^{15}-10^{18}\) eV.

As can be seen from Fig.~\ref{fig:ykt_45}a, the obtained spectrum has two features at the energy \(\sim 3\cdot 10^{15}\) eV (first knee) and at the energy \(\sim 10^{17}\) eV (second knee). The first knee is characterized by the slope \(\gamma_{1} = 2.70\pm0.03\) and \(\gamma_{2}=3.12\pm0.03\) and the second knee \(\gamma_{3}=2.92\pm0.03\) and \(\gamma_{4}=3.34\pm0.04\). 

Fig.~\ref{fig:ykt_45}b shows a comparison with other experiments: TALE~\cite{Abu-Zayyad2022395347}, TA BR / LR~\cite{Abbasi201680}, KASCADE-Grande~\cite{Apel201236183} and Tunka~\cite{Prosin201475694}. There is a good agreement of the spectra in the energy range \(10^{16}-10^{18}\) eV. All experiments have a break in the spectrum at \(\sim 10^{17}\) eV. The discrepancies in the spectrum is partly due to the different methods of energy estimation at various air shower experiments and to some extent by the different effective thresholds of the experiments themselves. Table~\ref{tab:tab_3} shows the comparison of spectrum slopes between different experiments. 

\begin{table}[h!]
	\caption{Comparison of the spectrum slopes of different experiments}
	\label{tab:tab_3}
	\begin{tabular}{|c|c|c|c|c|}
		\hline
		Energy & Yakutsk & Tunka & KASCADE-Grande & TALE \\ 
		\hline
		\(\Delta E\), eV & \(\gamma\pm stat\pm sys\) & \(\gamma\pm stat\pm sys\) & \(\gamma\pm stat\pm sys\) & \(\gamma\pm stat\) \\
		\hline
		\((1.2-5)\cdot 10^{15} \) & \(-2.7\pm 0.04\pm 0.10 \) & - & - & - \\
		\hline
		\((5-20)\cdot 10^{15} \)& \(-3.12\pm 0.03\pm 0.07 \)& \(-3.26\pm 0.01\pm 0.01 \) & - & \( -3.09\pm0.01\) \\
		\hline
		\((2-20) \cdot 10^{16} \) & \(-2.92\pm 0.03\pm 0.06 \) & \(-2.98\pm 0.01 \pm 0.01 \) & \(-2.95\pm 0.05\pm 0.02 \) & \(-2.89\pm 0.01 \) \\
		\hline
		\((2-30)\cdot 10^{17} \) & \(-3.34\pm 0.04\pm 0.05 \) & \(-3.35\pm 0.01\pm 0.01 \) & \(-3.24\pm 0.08\pm 0.05\)& \(-3.20\pm 0.02\) \\
		\hline
	\end{tabular}	
\end{table}

\section{Conclusion}

Using a large dataset of Cherenkov light measurements registered for the period 1994-2014 years, air shower energy is estimated by the energy balance method and the spectrum of cosmic rays in the energy range \(10^{16}-10^{18}\) eV is obtained. At the energy of \(\sim10^{17}\) eV spectrum slope changes from -2.92 to -3.34, which is associated with astrophysical processes in our galaxy, as well as with extragalactic processes. The “second knee” phenomenon can be explained as a transition from galactic to extragalactic cosmic rays.
\\

% TODO: include funding information
\paragraph{Funding information}
This work was carried out in the framework of research project No. AAAA-A21-121011990011-8 by the Ministry of Science and Higher Education of the Russian Federation.

\bibliography{bib_second_knee}
%\bibliographystyle{SciPost_bibstyle}

% TODO:
% Provide your bibliography here. You have two options:

% FIRST OPTION - write your entries here directly, following the example below, including Author(s), Title, Journal Ref. with year in parentheses at the end, followed by the DOI number.
%\begin{thebibliography}{99}
%\bibitem{1931_Bethe_ZP_71} H. A. Bethe, {\it Zur Theorie der Metalle. i. Eigenwerte und Eigenfunktionen der linearen Atomkette}, Zeit. f{\"u}r Phys. {\bf 71}, 205 (1931), \doi{10.1007\%2FBF01341708}.
%\bibitem{arXiv:1108.2700} P. Ginsparg, {\it It was twenty years ago today... }, \url{http://arxiv.org/abs/1108.2700}.
%\end{thebibliography}

% \bibliographystyle{SciPost_bibstyle} % Include this style file here only if you are not using our template

\nolinenumbers

\end{document}